\documentclass[english,two column]{revtex4}
\usepackage[T1]{fontenc}
\usepackage[latin9]{inputenc}
\usepackage{graphicx}
\usepackage{esint}

\makeatletter

\providecommand{\tabularnewline}{\\}

\@ifundefined{textcolor}{}
{%
 \definecolor{BLACK}{gray}{0}
 \definecolor{WHITE}{gray}{1}
 \definecolor{RED}{rgb}{1,0,0}
 \definecolor{GREEN}{rgb}{0,1,0}
 \definecolor{BLUE}{rgb}{0,0,1}
 \definecolor{CYAN}{cmyk}{1,0,0,0}
 \definecolor{MAGENTA}{cmyk}{0,1,0,0}
 \definecolor{YELLOW}{cmyk}{0,0,1,0}
}

\makeatother

\usepackage{babel}
\begin{document}

\title{Quantum Dephasing of Interacting Quantum Dot Induced by the Superconducting
Proximity Effect}

\author{Y. N. Fang$^{1,2}$, S. W. Li$^{2,4}$, L. C. Wang$^{3,4}$, and
C. P. Sun$^{1,2,4}$}

\email{cpsun@csrc.ac.cn}

\affiliation{$^{1}$State Key Laboratory of Theoretical Physics, Institute of
Theoretical Physics, Chinese Academy of Sciences, and University of
the Chinese Academy of Sciences, Beijing 100190, China\\
$^{2}$Synergetic Innovation Center of Quantum Information and Quantum
Physics, University of Science and Technology of China, Hefei, Anhui
230026, China\\
$^{3}$School of Physics and Optoelectronic Technology, Dalian University
of Technology, Dalian 116024, China\\
 $^{4}$Beijing Computational Science Research Center, Beijing 100084,
China\\
}
\begin{abstract}
The proximity effect (PE) between superconductor and confined electrons
can induce the effective pairing phenomena of electrons in nanowire
or quantum dot (QD). Through interpreting the PE as an exchange of
virtually quasi-excitation in a largely gapped superconductor, we
found that there exists another induced dynamic process. Unlike the
effective pairing that mixes the QD electron states coherently, this
extra process leads to dephasing of the QD. In a case study, the dephasing
time is inversely proportional to the Coulomb interaction strength
between two electrons in the QD. Further theoretical investigations
imply that this dephasing effect can decrease the quality of the zero
temperature mesoscopic electron transportation measurements by lowering
and broadening the corresponding differential conductance peaks.
\end{abstract}
\maketitle

\section{Introduction}

Superconducting proximity effect (PE) was originally explored in the
study of the normal-superconductor metals interface \cite{First proximity paper}.
When a sub-gap electron was injected from the normal side to a highly
transparent interface, the Andreev reflection can happen, such that
a hole is retro-reflected out from the superconducting side \cite{Andreev 1,Andreev 2}.
This process is energy-favorable because Cooper pairs condense in
the BCS ground state. As a result the extra electron pairs injected
from the normal side can be absorbed without perturbing the superconductor.

Owing to the fast development in nanofabrication technology, novel
effects in physics and related phenomena that discovered in hybrid
devices have attracted much research interests especially in the searching
for Majorana Fermions. Many previous investigations had taken the
advantages of PE in producing exotic superconductivity with a p-wave
component \cite{PE proposal 1,PE proposal 2,PE proposal 3,PE proposal 4}.
In those proposals, usually a nanowire with strong spin-orbit coupling
is placed in close contact with a bulk s-wave pairing superconductor
(SC), such that electron can tunnel between SC and the nanowire. If
one focus on physics inside the SC gap, an effective Hamiltonian for
the nanowire can be derived with additional terms describing electron
pair creation or annihilation \cite{Proximity induced pairing 1,Proximity induced pairing 2},
which are referred as PE induced pairing terms. Although those proposals
stimulate a series of experimental works with significant results
\cite{nanowire experiment 1,nanowire experiment 2,nanowire experiment 3},
the nanowire-based setup has its drawbacks such as difficulties in
manipulating chemical potential \cite{nanowire experiment 1,QD chain 1}
as well as fragile of the induced pairing potential against disorder
\cite{QD chain 2}. 

To overcome those obstacles, analogy in other systems \cite{Other modified nanowire prop.}
or modified solid-system proposals have been suggested. A notable
trend among those is to replace the nanowire by a chain of coupled
quantum dots (QD), which is introduced by J. D. Sau \textit{et al}
in ref. \cite{QD chain 1} and further developed by many groups \cite{QD chain 2,QD chain 4,QD chain 3,QD chain 5}.
However, many discussions in this aspect adopt directly the discretized
version of the previous effective Hamiltonians of the nanowire that
are obtained based on the single electron picture. Meanwhile, in QD
related transport studies, there are also researches which point out
that a large Coulomb repulsion can suppress onsite PE pairing \cite{Coulomb int kill PE 1,single-level anderson 1},
and the above is actually the key idea of the Cooper pair beam splitter
\cite{Proximity induced pairing 2}. Those observations motivated
us to investigate whether interaction effect can be important or not
in discussing PE on the QD.

\begin{figure}
\begin{centering}
\includegraphics[scale=0.43]{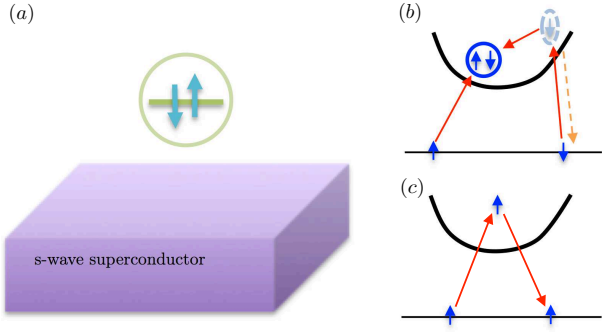}
\par\end{centering}

\caption{(Color online) (a) Schematic plot of a single level quantum dot (QD)
in proximity to a bulk s-wave pairing superconductor (SC). (b) Pictorial
representation of the Andreev reflection, where the black parabolic
curve denotes the quasi-excitation state of the SC and the horizontal
line at the bottom denotes energy of the QD level. (c) Pictorial illustration
on virtual process that is relevant for the proximity induced dephasing
effect.}
\end{figure}
In this paper, we present a theoretical approach that can be used
to explore PE on the QD system when the onsite Coulomb interaction
strength was much smaller than the SC gap. The idea is based on adiabatic
eliminating the SC quasi-excitation, since the electron tunneling
between QD and SC becomes virtual due to the large SC gap. Then, we
apply this approach to a simplified model with only one QD. The corresponding
effective Hamiltonian of QD reduces to previous ones if the Coulomb
interaction is ignored. However, when the Coulomb interaction was
included, new terms representing interactions between QD and SC are
identified besides the previously obtained pairing terms. To the leading
order, the main contribution comes from a dephasing term that can
force the QD to evolve into a mixed state.

This paper is organized as follows, in Sec. II we bring in the QD
model and use the adiabatic elimination scheme to derive the effective
Hamiltonian of QD. In Sec. III, the decoherence factor as well as
the dephasing time of QD are evaluated by employing a semi-classical
treatment. Physical implication of the PE induced dephasing is discussed
in Sec. IV by studying the transport properties of QD in a three-terminals
quantum point contact device. Finally a summary is given in Sec. V.

\section{Adiabatic elimination and proximity induced dephasing}

Let us consider a QD which is located near a bulk SC, as schematically
shown in Fig.1(a). For simplicity consideration we model the QD as
a single band Anderson impurity \cite{single-level anderson 1,single-level anderson 2,single-level anderson 3},
although generalization to the cases of finite size \cite{non-Anderson impurtity for QD,non-Anderson impurtity for QD 2}
as well as multiple levels \cite{multi-level QD 1,multi-level QD 2}
are possible. The SC is described by an s-wave pairing mean-field
BCS Hamiltonian with real pairing potential $\Delta$. Consequently,
Hamiltonian of the QD-SC hybrid system is written as $H=H_{d}+H_{s}+H_{sd},$
with
\begin{equation}
H_{d}=\varepsilon\sum_{\sigma}n_{\sigma}+Un_{\uparrow}n_{\downarrow},\label{Hd}
\end{equation}
\begin{equation}
H_{s}=\sum_{\mathbf{k}}[\xi_{\mathbf{k}}(c_{k}^{\dagger}c_{k}+c_{-k}^{\dagger}c_{-k})+\Delta(c_{k}^{\dagger}c_{-k}^{\dagger}+c_{-k}c_{k})],\label{Hs}
\end{equation}
and
\begin{equation}
H_{sd}=\sum_{\mathbf{k},\sigma}(t_{\mathbf{k}}d_{\sigma}^{\dagger}c_{\mathbf{k}\sigma}+h.c.).\label{Hsd}
\end{equation}

Here, $H_{d}$ and $H_{s}$ are the Hamiltonians of the QD and the
SC, respectively. $n_{\sigma}=d_{\sigma}^{\dagger}d_{\sigma}$ is
QD electron number operator with electron spin $\sigma\in\{\uparrow,\downarrow\}$.
$\varepsilon=\epsilon_{d}-\mu$ is energy of the QD level measured
from the chemical potential $\mu$ of the SC, $U$ is the onsite Coulomb
interaction strength for double occupation of the QD. $\xi_{\mathbf{k}}=\mathbf{k}^{2}/(2m)-\mu$
is the kinetic energy of a free electron measured from $\mu$. We
use $k=\{\mathbf{k},\uparrow\}$ and $-k=\{-\mathbf{k},\downarrow\}$
to jointly label momentum and spin for the SC electrons. $H_{sd}$
describes the single electron tunneling process. For tunneling that
happens locally in space, as relevant to quantum point contact (QPC),
the corresponding tunneling probability amplitude $t_{\mathbf{k}}=t_{0}\exp\{i\mathbf{k}\cdot\vec{r}\}$.
Here $t_{0}$ is assumed to be real and $\vec{r}$ denotes the location
of QD.

$H_{s}$ can be diagonalized by the following Bogoliubov transformation
\begin{equation}
\gamma_{k}=\cos\theta_{k}c_{k}+\sin\theta_{k}c_{-k}^{\dagger},\mbox{ }\gamma_{-k}=\cos\theta_{k}c_{-k}-\sin\theta_{k}c_{k}^{\dagger},\label{Bog trans}
\end{equation}
where $\tan2\theta_{k}=\Delta/\xi_{\mathbf{k}}$ and notice that $\{\gamma_{k},\gamma_{-k'}^{\dagger}\}=0$.
After the Bogoliubov transformation, $H_{s}$ is rewritten as
\begin{equation}
H_{s}=\sum_{\mathbf{k}}E_{\mathbf{k}}(\gamma_{k}^{\dagger}\gamma_{k}+\gamma_{-k}^{\dagger}\gamma_{-k}),\label{BCS H dia}
\end{equation}
where $E_{\mathbf{k}}=\sqrt{\xi_{\mathbf{k}}^{2}+\Delta^{2}}$ is
the elementary excitation spectrum of the SC. $H_{sd}$ is also rewritten
as follows in terms of quasi-particle operators $\gamma_{k}$ and
$\gamma_{-k}$, i.e.,
\begin{equation}
H_{sd}=\sum_{\mathbf{k}}[\eta_{\mathbf{k}}(d_{\uparrow}^{\dagger}\gamma_{k}-d_{\downarrow}\gamma_{-k}^{\dagger})-\lambda_{\mathbf{k}}(d_{\downarrow}\gamma_{k}+d_{\uparrow}^{\dagger}\gamma_{-k}^{\dagger})+h.c.].\label{Hsd BCS dia}
\end{equation}

In the sub-gap regime (SGR), i.e., both $U\ll\Delta$ and $\varepsilon\ll\Delta$,
all relevant QD states are within the SC gap. Thus the minimal energy
gap between electron states in the QD and quasi-excitation states
in the SC is $\delta\omega=\min\{\left|\Delta-\varepsilon\right|,\left|\Delta-\varepsilon-U\right|\}$.
Suppose that the tunneling strength between SC and QD further satisfies
$t_{0}/\delta\omega\ll1$, then one can eliminated $H_{sd}$ up to
the first order by performing the canonical transformation $\exp(S)$
\cite{Adiabatic elimination 1,Adiabatic elimination 2}, where
\begin{equation}
S=\sum_{\mathbf{k}}[\alpha_{\mathbf{k}}(D_{\uparrow}^{\dagger}\gamma_{k}+D_{\downarrow}\gamma_{-k}^{\dagger})-\beta_{\mathbf{k}}(\tilde{D}_{\downarrow}\gamma_{k}-\tilde{D}_{\uparrow}^{\dagger}\gamma_{-k}^{\dagger})-h.c.].\label{S}
\end{equation}
Here, $\alpha_{\mathbf{k}}$, $\beta_{\mathbf{k}}$, $D_{\sigma}$
and $\tilde{D}_{\sigma}$ are
\begin{equation}
\alpha_{\mathbf{k}}=\frac{\eta_{\mathbf{k}}}{E_{\mathbf{k}}-\varepsilon},\mbox{ }\beta_{\mathbf{k}}=\frac{\lambda_{\mathbf{k}}}{E_{\mathbf{k}}+\varepsilon},\label{S coef 1}
\end{equation}
and
\begin{equation}
D_{\sigma}=d_{\sigma}(1+\frac{Un_{\bar{\sigma}}}{E_{\mathbf{k}}-\varepsilon-U}),\mbox{ }\tilde{D}_{\sigma}=d_{\sigma}(1-\frac{Un_{\bar{\sigma}}}{E_{\mathbf{k}}+\varepsilon+U}).\label{S coef 2}
\end{equation}
Here, $\eta_{\mathbf{k}}=t_{\mathbf{k}}\cos\theta_{k}$ and $\lambda_{\mathbf{k}}=t_{\mathbf{k}}\sin\theta_{k}$.
$\bar{\sigma}$ denotes the spin direction opposite to $\sigma$.

After the canonical transformation, Hamiltonian of the hybrid system
is written as
\begin{equation}
e^{-S}He^{S}\approx H_{d}^{\mathrm{eff}}+H_{s}^{\mathrm{eff}}+H_{sd}^{\mathrm{eff}},\label{trans H}
\end{equation}
where $H_{d}^{\mathrm{eff}}$ is given by
\begin{equation}
H_{d}^{\mathrm{eff}}=H_{d}+(\Delta_{d}d_{\uparrow}^{\dagger}d_{\downarrow}^{\dagger}+h.c.),\label{Hd_eff}
\end{equation}
$\Delta_{d}$ is PE induced pairing potential in the QD, 
\begin{equation}
\Delta_{d}=t_{0}^{2}\sum_{\mathbf{k}}\frac{\Delta}{E_{\mathbf{k}}^{2}-\varepsilon^{2}}=\frac{\pi t_{0}^{2}N_{0}\Delta}{\sqrt{\Delta^{2}-\varepsilon^{2}}}.\label{Dd}
\end{equation}
Here, $N_{0}$ is normal state density of state (DOS) at $\mu$. Eq.(\ref{Dd})
has been obtained before in weak tunneling limit \cite{PE induced pairing in waek tunneling 1,Proximity induced pairing 1}.
Notice that $\Delta_{d}$ saturates to a constant value that independent
from $\Delta$ in the SGR.

Notice that generally all parameters involved in $H_{d}$ of Eq.(\ref{Hd_eff})
should be renormalized when compared with parameters defined in Eq.(\ref{Hd}),
but in the following we shall not distinguish this difference and
still adopt the previous notations for an isolated QD. Furthermore,
we take $H_{s}^{\mathrm{eff}}\approx H_{s}$ since the back action
of the QD on the SC should be minor.

$H_{sd}^{\mathrm{eff}}$ denotes interactions between QD and SC, full
expression of this term is given in the Appendix A. By averaging $H_{sd}^{\mathrm{eff}}$
over the BCS ground state of the SC, the only contributing term has
the following form
\begin{equation}
H_{sd}^{\mathrm{eff}}\approx\sum_{\sigma}n_{\sigma}\otimes R_{\sigma},\label{PID}
\end{equation}
where $R_{\sigma}$ are two operators acting only on Hilbert space
of the SC, i.e.,
\begin{eqnarray}
R_{\uparrow} & = & \frac{1}{2}\sum_{\mathbf{k},\mathbf{k}'}[\tilde{\alpha}_{\mathbf{k}'}(\eta_{\mathbf{k}}^{*}\eta_{\mathbf{k}'}\gamma_{-k'}^{\dagger}\gamma_{-k}-\lambda_{\mathbf{k}}^{*}\eta_{\mathbf{k}'}\gamma_{k}^{\dagger}\gamma_{-k'}^{\dagger})\nonumber \\
 &  & +\tilde{\beta}_{\mathbf{k}'}(\lambda_{\mathbf{k}}^{*}\lambda_{\mathbf{k}'}\gamma_{k'}\gamma_{k}^{\dagger}+\eta_{\mathbf{k}}^{*}\lambda_{\mathbf{k}'}\gamma_{k'}\gamma_{-k})+h.c.],\label{R_uparrow}
\end{eqnarray}
and
\begin{eqnarray}
R_{\downarrow} & = & \frac{1}{2}\sum_{\mathbf{k},\mathbf{k}'}[-\tilde{\beta}_{\mathbf{k}'}(\lambda_{\mathbf{k}}^{*}\lambda_{\mathbf{k}'}\gamma_{-k}\gamma_{-k'}^{\dagger}+\eta_{\mathbf{k}}^{*}\lambda_{\mathbf{k}'}\gamma_{-k'}^{\dagger}\gamma_{k}^{\dagger})\nonumber \\
 &  & +\tilde{\alpha}_{\mathbf{k}'}(\eta_{\mathbf{k}}^{*}\eta_{\mathbf{k}'}\gamma_{k}^{\dagger}\gamma_{k'}-\lambda_{\mathbf{k}}^{*}\eta_{\mathbf{k}'}\gamma_{-k}\gamma_{k'})+h.c.],\label{R_downarrow}
\end{eqnarray}
where 
\begin{equation}
\tilde{\alpha}_{\mathbf{k}}=\frac{U}{(E_{\mathbf{k}}-\varepsilon)(E_{\mathbf{k}}-\varepsilon-U)},\mbox{ }\tilde{\beta}_{\mathbf{k}}=\frac{U}{(E_{\mathbf{k}}+\varepsilon)(E_{\mathbf{k}}+\varepsilon+U)}.\label{para R_sig}
\end{equation}
Notice that both $\tilde{\alpha}_{\mathbf{k}}$ and $\tilde{\beta}_{\mathbf{k}}$
approach to zero as $U$ goes to zero, thus indicating the Coulomb
interaction is necessary in producing coupling term given by Eq.(\ref{PID}).

A basis for the QD is chosen as $\{|0\rangle,|\sigma\rangle,|d\rangle\}$,
where $|0\rangle$ and $|d\rangle$ denote the emptily as well as
the doubly occupied state of the QD, and $|\sigma\rangle$ are single
occupation states with spin $\sigma$. It follows from Eq.(\ref{Hd_eff})
that the PE induced pairing term causes mixing of basis states in
the subspace spanned by $|0\rangle$ and $|d\rangle$, while the single
electron subspace $\{|\sigma\rangle\}$ remains unaffected by this
term. However, the quantum coherence between spin-up and spin-down
states in the latter subspace can still lose due to coupling with
the SC through the term $H_{sd}^{\mathrm{eff}}$. This is because,
with the additional term $H_{sd}^{\mathrm{eff}}$, the superconductor
now can evolve differently depending on spin orientation of the QD
electron. This PE induced dephasing (PID) effect would be absent if
one ignores the onsite Coulomb interaction.

\section{the dephasing time}

The PID process and subsequent decoherence of the QD deserve more
detailed investigation, since quantum coherence is a crucial resource
in implementing various quantum computations \cite{quantum coherence 1,quantum coherence 2}.
To this end, we estimate characteristic dephasing time of the QD by
studying its decoherence factor.

Since the single electron subspace is decoupled from the subspace
$\{|0\rangle,|d\rangle\}$, we shall restrict following discussion
only to the case of single electron. Consider following initial state
for the hybrid system with one electron in the QD
\begin{equation}
|\Phi(0)\rangle=\sum_{\sigma}w_{\sigma}|\sigma\rangle\otimes|BCS\rangle,\label{PID initial state}
\end{equation}
where $|w_{\sigma}|^{2}$ is probability for electron to occupy the
state $|\sigma\rangle$. $|BCS\rangle$ denotes the BCS ground state
of the SC. After a period of evolution, the final state becomes
\begin{equation}
|\Phi(t)\rangle=\sum_{\sigma}w_{\sigma}e^{-is[\sigma]\varepsilon t}|\sigma\rangle\otimes e^{-ih_{\sigma}t}|BCS\rangle.\label{PID final state}
\end{equation}
Here, $s[\uparrow]=1$ and $s[\downarrow]=-1$. $h_{\sigma}$ is defined
as $h_{\sigma}=H_{s}+R_{\sigma}$.

To discuss the QD dephasing, we consider the reduced density matrix
of the QD system. This is given by
\begin{eqnarray}
\rho(t) & = & \mbox{tr}_{\mathrm{sc}}\{|\Phi(t)\rangle\langle\Phi(t)|\}\nonumber \\
 & = & |c_{\uparrow}|^{2}|\uparrow\rangle\langle\uparrow|+|c_{\downarrow}|^{2}|\downarrow\rangle\langle\downarrow|\nonumber \\
 &  & +(c_{\uparrow}c_{\downarrow}^{*}e^{-2i\varepsilon t}|\uparrow\rangle\langle\downarrow|D(t)+h.c.).\label{RDM of QD}
\end{eqnarray}
Here, $\mathrm{tr}_{\mathrm{sc}}$ means tracing over the degree of
freedom of the SC. $D(t)$ is the decoherence factor of QD\cite{decoherence factor},
\begin{equation}
D(t)=\langle e^{ih_{\downarrow}t}e^{-ih_{\uparrow}t}\rangle_{\mathrm{sc}},\label{Decoherence factor}
\end{equation}
where the averaging is taken with respect to $|BCS\rangle$.

Since $h_{\sigma}$ is quadratic in quasi-particle operators, the
exact evaluation of $D(t)$ is always possible but too complicated
to be done. Therefore, we adopt a semi-classical treatment to estimate
$D(t)$. We remark that this method is equivalent to second order
cumulant expansion \cite{second order cum} if sum over $\mathbf{k}\ne\mathbf{k}'$
terms involved in $R_{\sigma}$ are ignored. The idea of this semi-classical
method is to replace $R_{\sigma}$ by a random number $\mathcal{R}_{\sigma}$,
whose mean value and covariance are determined quantum mechanically
by using following relations
\begin{equation}
\langle\mathcal{R}_{\sigma}\rangle_{\mathrm{cl}}=\langle R_{\sigma}\rangle_{\mathrm{sc}},\mbox{ }\langle\langle\mathcal{R}_{\sigma}\mathcal{R}_{\sigma'}\rangle\rangle_{\mathrm{cl}}=\langle R_{\sigma}R_{\sigma'}\rangle_{\mathrm{sc}}-\langle R_{\sigma}\rangle_{\mathrm{sc}}\langle R_{\sigma'}\rangle_{\mathrm{sc}},\label{semi-classic app}
\end{equation}
Also the quantum mechanically trace in $D(t)$ is replaced by averaging
over probability density function (PDF) of $\mathcal{R}_{\sigma}$. 

In this way, $D(t)$ is rewritten as
\begin{equation}
D(t)\approx\langle e^{-i(\mathcal{R}_{\uparrow}-\mathcal{R}_{\downarrow})t}\rangle_{\mathrm{cl}}.\label{Decoherence factor sc app}
\end{equation}
Using the following cumulant expansion formula \cite{Van Kampen book}
\begin{eqnarray}
 &  & \langle e^{\alpha\delta x+\beta\delta y}\rangle_{\mathrm{cl}}\nonumber \\
 & \approx & e^{\alpha\langle\delta x\rangle_{\mathrm{cl}}+\beta\langle\delta y\rangle_{\mathrm{cl}}+\frac{1}{2}\alpha^{2}\langle\langle\delta x^{2}\rangle\rangle_{\mathrm{cl}}+\frac{1}{2}\beta^{2}\langle\langle\delta y^{2}\rangle\rangle_{\mathrm{cl}}+\alpha\beta\langle\langle\delta x\delta y\rangle\rangle_{\mathrm{cl}}},\nonumber \\
\label{cumulant expan}
\end{eqnarray}
$D(t)$ is further expressed as
\begin{equation}
D(t)\approx e^{-i\Omega t}e^{-\Gamma^{2}t^{2}},\label{decoherence factor sc app result}
\end{equation}
where
\begin{equation}
\Omega=2\langle\mathcal{R}_{\uparrow}\rangle_{\mathrm{cl}},\mbox{ }\Gamma=\sqrt{\frac{1}{2}\sum_{\sigma}\langle\langle\mathcal{R}_{\sigma}^{2}\rangle\rangle_{\mathrm{cl}}-\langle\langle\mathcal{R}_{\uparrow}\mathcal{R}_{\downarrow}\rangle\rangle_{\mathrm{cl}}}.\label{decoherence factor sc app para}
\end{equation}
Notice that in deriving Eq.(\ref{decoherence factor sc app para}),
we have used $\langle\mathcal{R}_{\sigma}+\mathcal{R}_{\bar{\sigma}}\rangle_{\mathrm{cl}}=0$.
Proof of this relation can be found in Appendix B.

According to Eqs.(\ref{decoherence factor sc app result},\ref{decoherence factor sc app para}),
the dephasing time of QD is defined as $T_{\mathrm{ph}}=\Gamma^{-1}$.
If QD energy is chosen such that $\varepsilon=-U/2$, then $T_{\mathrm{ph}}$
has the following analytical expression (see Appendix B), i.e.,
\begin{equation}
T_{\mathrm{ph}}=|\zeta(\Delta)|\frac{\Delta}{U\Delta_{d}},\label{dephasing time}
\end{equation}
where $\zeta^{2}(\Delta)=\pi\sqrt{(\Delta^{2}-U^{2}/4)/(\Lambda_{D}^{2}-\Delta^{2})}$
depends mainly on $\Delta$ since $U\ll\Delta$. $\Lambda_{D}$ is
characteristic width of energy shell where the SC electron-electron
effective attraction is non-zero. It follows from Eq.(\ref{Dd}) that
$\Delta_{d}$ is asymptotically independent from $\Delta$ in the
SGR. Then $T_{\mathrm{ph}}$ is proportional to $\Delta$ and inversely
proportional to the Coulomb interaction strength, which means that
the dephasing effect becomes weaker as SC gap became larger or Coulomb
interaction became smaller.

\section{Observable effects of the proximity induced dephasing}

The PID effect manifest itself as a quantum fluctuation on QD levels
as a result of virtual quasi-particle exchanging with the SC. This
modulation on the energy level can be probed by the measurement on
mesoscopic transport through a QPC, which at low temperature provides
the information about the local density of states at the QD \cite{STM prob local density state}.
Besides this, although it has been predicted that an observation of
a quantized zero-bias peak in differential conductance measurement
can be regarded as a necessary condition for the judgment on the existence
of Majorana quasi-particles \cite{quantise ZBP theoretic}, the experimental
results on transport studies display peaks that are much lower than
this theoretical prediction \cite{nanowire experiment 1,ZBP less then 2 e2/h}.
This fact has motivated several explorations on the possible explanations
\cite{ZBP explain 1,ZBP explain 2,ZBP explain 3}, which also attract
us to consider the PID effect on transport setups. In fact, there
has been some transport based researches on the effect of dephasing
caused by electron-electron interaction in QD systems \cite{QD dephasing,QD dephasing 2}.

Let us consider a three terminals transport device with two leads
in normal and one in the superconducting states, as schematically
shown in Fig.2(a). Similar multi-terminals devices have been investigated
in ref. \cite{QD dephasing,multi-terminal 1,multi-terminal 2}. The
Hamiltonian of the system is written as $H_{\mathrm{tot}}=H+\sum_{\nu=L,R}(H_{\nu}+H_{\nu d})$,
with
\begin{equation}
H_{\nu}=\sum_{\mathbf{k},\sigma}\xi_{\mathbf{k}}c_{\nu\mathbf{k}\sigma}^{\dagger}c_{\nu\mathbf{k}\sigma},\label{H_leads}
\end{equation}
and
\begin{equation}
H_{\nu d}=\sum_{\mathbf{k},\sigma}(t_{\mathbf{k}}^{(\nu)}d_{\sigma}^{\dagger}c_{\nu\mathbf{k}\sigma}+h.c.).\mbox{ }(\nu=L,R)\label{H_tunel}
\end{equation}
Here $H$ is given by Eqs.(\ref{Hd}-\ref{Hsd}), which describes
the SC lead, the QD, as well as single electron tunneling that happened
between the two. $H_{\nu}$ is Hamiltonian of the normal lead $\nu$
with chemical potential $\mu_{\nu}$, notice that the kinetic energy
is again measured from the SC chemical potential $\mu$. $H_{\nu d}$
denotes the electron tunneling between the QD and normal lead $\nu$.
For the localized tunneling $t_{\mathbf{k}}^{(\nu)}=t_{\nu}\exp\{i\mathbf{k}\cdot\vec{r}\}$
and $t_{\nu}$ are assumed to be real.

Suppose that the pairing potential of the SC lead is chosen to satisfy
the SGR conditions. Then according to the results in Sec. II, $H$
can be replaced by $H_{d}^{\mathrm{eff}}+H_{sd}^{\mathrm{eff}}$,
i.e., the role of the SC lead is included effectively by using the
adiabatic elimination approach. Therefore, in this case we can focus
on the current through the QD between two normal leads. The current
from lead $\nu$ is written as
\begin{eqnarray}
I_{\nu} & = & \langle\frac{d}{dt}N_{\nu}\rangle=-ie\langle[H_{\mathrm{tot}},N_{\nu}]\rangle.\label{current_def}
\end{eqnarray}
Here, $N_{\nu}=\sum_{\mathbf{k},\sigma}c_{\nu\mathbf{k}\sigma}^{\dagger}c_{\nu\mathbf{k}\sigma}$
is the total number of electrons in the lead $\nu$. In the Heisenberg
picture, the above averaging is taken with respect to initial state
of the whole system. Notice that the chemical potentials of two normal
leads are not always the same due to applied bias voltage, as shown
in Fig.2(b).

\begin{figure}
\begin{centering}
\includegraphics[scale=0.25]{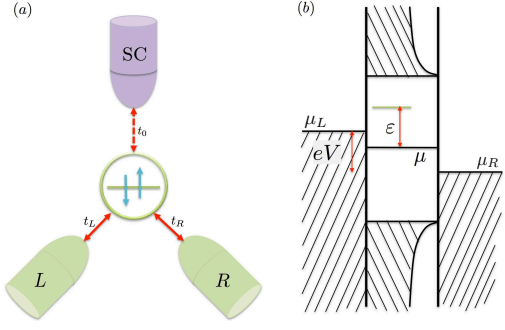}
\par\end{centering}

\caption{(Color online) (a) Schematic of a three-terminals device in quantum
point contact (QPC) with a single level quantum dot (QD), one of those
leads is in the BCS superconducting ground state (shown in purple).
In the sub-gap regime (SGR), real single electron tunneling between
QD and the superconducting (SC) lead does not happen, as stressed
by dashed red arrows. (b) Energy landscape of the multi-terminals
QPC system, where the bias voltage is applied between two normal leads.
The QD level with energy $\varepsilon$ (measured from SC chemical
potential $\mu$) is shown by green solid line inside the SC gap.}
\end{figure}
The total current through the QD is given by $I_{\mathrm{tot}}=(I_{L}-I_{R})/2$.
In steady state, $I_{\mathrm{tot}}$ can be expressed in terms of
Green's function of the QD by using the non-equilibrium techniques
\cite{Wingreen}, i.e.,
\begin{equation}
I_{\mathrm{tot}}=-\frac{e\Gamma_{0}}{h}\sum_{\sigma}\int d\omega\mathrm{Im}[g_{\sigma}^{r}(\omega)][f_{L}(\omega+\mu)-f_{R}(\omega+\mu)],\label{I_tot}
\end{equation}
where the Plank constant $h$ is written out explicitly. The above
equation is valid in the width band limit \cite{Wingreen}, where
the DOS of the normal leads are assumed as constant within the energy
spectrum of the single level QD. We also assumed that $t_{L}=t_{R}=t_{1}$,
thus the line width function $\Gamma_{0}=2\pi t_{1}^{2}N_{0}$ is
the same for both normal leads \cite{multi-level QD 1}. $f_{\nu}(\omega)$
is the equilibrium state Fermi-distribution at lead $\nu$, i.e.,
$f_{\nu}(\omega)=\{\exp[\beta(\omega-\mu_{\nu})]+1\}^{-1}$.

$g_{\sigma}^{r}(\omega)$ is Fourier transform of the retarded QD
Green's function,
\begin{equation}
g_{\sigma}^{r}(t,t')=-i\theta(t-t')\langle\{d_{\sigma}(t),d_{\sigma}^{\dagger}(t')\}\rangle,\label{g_retard}
\end{equation}
which is the key quantity in evaluating $I_{\mathrm{tot}}$. 

At low temperature, the Fermi-distribution function can be approximated
by the step-function, i.e., $f_{\nu}(\omega)\approx\theta(\mu_{\nu}-\omega)$.
Suppose $\mu_{R}=\mu$, then $I_{\mathrm{tot}}$ is rewritten as
\begin{equation}
I_{\mathrm{tot}}=-\frac{e\Gamma_{0}}{h}\sum_{\sigma}\int_{0}^{eV}d\omega\mathrm{Im}g_{\sigma}^{r}(\omega),\label{I_tot at low temp}
\end{equation}
where $V=(\mu_{L}-\mu_{R})/e$ denotes the bias voltage across two
normal leads. Differential conductance is then given by
\begin{equation}
\frac{dI_{\mathrm{tot}}}{dV}=-\frac{e^{2}}{h}\Gamma_{0}\sum_{\sigma}\mathrm{Im}g_{\sigma}^{r}(eV).\label{dI/dV}
\end{equation}

According to the derivation detailed in Appendix C, at the mean-field
level \cite{MF app 1,MF app 2} $g_{\sigma}^{r}(\omega)$ is calculated
by using the motion equation method, i.e.,
\begin{equation}
g_{\sigma}^{r}(\omega)=\frac{D_{h}(\omega)G_{\sigma}^{(e)}(\omega)}{D_{e}(\omega)D_{h}(\omega)-\Delta_{d}^{2}G_{\bar{\sigma}}^{(e)}(\omega)G_{\sigma}^{(h)}(\omega)}\label{g_retard mean-field}
\end{equation}
with
\begin{equation}
D_{e}(\omega)=\omega-\varepsilon-(\mathcal{R}_{\sigma}-i\Gamma_{0})G_{\sigma}^{(e)}(\omega),\label{g_retard mean-field para 1}
\end{equation}
\begin{equation}
D_{h}(\omega)=\omega+\varepsilon+(\mathcal{R}_{\bar{\sigma}}+i\Gamma_{0})G_{\bar{\sigma}}^{(h)}(\omega),\label{g_retard mean-field para 2}
\end{equation}
\begin{equation}
G_{\sigma}^{(e)}(\omega)=1+\frac{U\langle n_{\bar{\sigma}}\rangle_{0}}{\omega-\varepsilon-U+i0^{+}},\label{g_retard mean-field para 3}
\end{equation}
and
\begin{equation}
\mbox{ }G_{\sigma}^{(h)}(\omega)=1-\frac{U\langle n_{\bar{\sigma}}\rangle_{0}}{\omega+\varepsilon+U+i0^{+}}.\label{g_retard mean-field para 4}
\end{equation}

Notice that in obtaining the above results, the semi-classical treatment
for the PID term introduced before in Sec. III had been used. Sub-index
in $\langle n_{\sigma}\rangle_{0}$ indicates to take the atomic limit
(AL) \cite{STM prob local density state}, where $t_{0}$ and $t_{1}$
are set to zero in evaluating the averaging $\left\langle ...\right\rangle _{0}$.

Due to the semi-classical treatment, Eq.(\ref{dI/dV}) should be averaged
over PDF of $\mathcal{R}_{\sigma}$,
\begin{equation}
\langle\frac{dI_{\mathrm{tot}}}{dV}\rangle_{\mathrm{cl}}=-\frac{e^{2}}{h}\Gamma_{0}\sum_{\sigma}\mathrm{Im}\langle g_{\sigma}^{r}(eV)\rangle_{\mathrm{cl}},\label{dI/dV semi-classical}
\end{equation}
Notice that in writing $\langle\mathrm{Im}g_{\sigma}^{r}(eV)\rangle_{\mathrm{cl}}=\mathrm{Im}\langle g_{\sigma}^{r}(eV)\rangle_{\mathrm{cl}}$,
we have implicitly assumed that there exists real PDF for $\mathcal{R}_{\sigma}$
that satisfy Eq.(\ref{semi-classic app}), but this point has not
been explicitly checked.

To calculate differential conductance we need to evaluate $\langle g_{\sigma}^{r}(\omega)\rangle_{\mathrm{cl}}$.
This can be done analytically in a special case where $\varepsilon=-U/2$
and $\langle n_{\sigma}\rangle_{0}=1/2$. As outlined in Appendix
D, in this case the Green's function can be evaluated by using a cumulant
expansion method similar to that employed in Sec. III. The result
is
\begin{eqnarray}
 &  & \langle g_{\sigma}^{r}(\omega)\rangle_{\mathrm{cl}}\nonumber \\
 & = & \frac{\omega}{\omega^{2}-\varepsilon^{2}+q_{\sigma}^{(e)}(\omega)+i\Gamma_{\sigma}\omega-\frac{\Delta_{d}^{2}\omega^{2}}{\omega^{2}-\varepsilon^{2}+q_{\bar{\sigma}}^{(h)}(\omega)+i\Gamma_{\bar{\sigma}}\omega}}.\nonumber \\
\label{g_retard semi-classical}
\end{eqnarray}
Here, $q_{\sigma}^{(e,h)}(\omega)$ depends on $\langle\mathcal{R}_{\sigma}\rangle_{\mathrm{cl}}$
and slightly shifts poles of $\langle g_{\sigma}^{r}(\omega)\rangle_{\mathrm{cl}}$,
their expressions are given by Eqs.(D7,D8). $\Gamma_{\sigma}$ is
related to differential conductance in the absence of the PID effect,
i.e.,
\begin{equation}
\Gamma_{\sigma}=\Gamma_{0}+\frac{h}{2\Gamma_{0}e^{2}}\frac{dI_{\mathrm{tot}}}{dV}|_{\mathcal{R_{\sigma}}\equiv0}\langle\langle\mathcal{R}_{\sigma}^{2}\rangle\rangle_{\mathrm{cl}}.\label{Gamma_sig corrected}
\end{equation}

\begin{figure}
\begin{centering}
\includegraphics[scale=0.218]{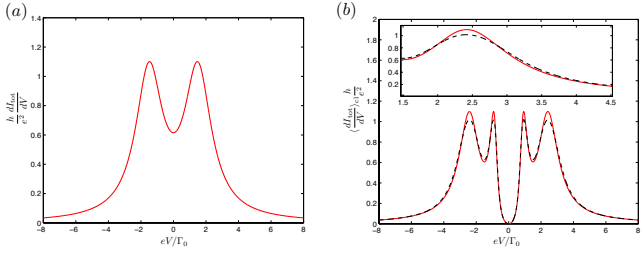}
\par\end{centering}

\caption{(Color online) Differential conductance calculated based on the semi-classical
treatment, black dashed (red solid) line is result with (without)
the proximity induced dephasing (PID) effect. We have set $\mu_{R}=\mu$
in those calculations, other parameters used are: $U/\Gamma_{0}=0\mbox{ and }3$
for (a) and (b), respectively; $\Gamma_{s}\equiv2\pi t_{0}^{2}N_{0}=3\Gamma_{0}$,
$\varepsilon=-U/2$, $\Delta/\Gamma_{0}=10$, $\Lambda_{D}/\Gamma_{0}=100$.
With this choice of parameters, the PE induced pairing potential $\Delta_{d}\sim\Delta/7$. }
\end{figure}
Since the differential conductance is related to the imaginary part
of the QD Green's function, consequently $\Gamma_{\sigma}$ determines
the width of conductance peaks. Therefore, the second term in Eq.(\ref{Gamma_sig corrected})
represents the PID effect on the peaks width. Notice that $(dI_{\mathrm{tot}}/dV)|_{\mathcal{R_{\sigma}}\equiv0}$
is positive. This can be checked by setting $q_{\sigma}^{(e,h)}(\omega)=0$
and $\Gamma_{\sigma}=\Gamma_{0}$ in Eq.(\ref{g_retard semi-classical})
followed by taking imaginary part. On the other hand, if $\Lambda_{D}\gg\Delta$,
it follows from the results shown in the Appendix B (see Eqs.(B10-B13),
for example) that $\langle\langle\mathcal{R}_{\sigma}^{2}\rangle\rangle_{\mathrm{cl}}$
are positive. As a result $\Gamma_{\sigma}\geq\Gamma_{0}$, this means
that due to the PID effect, differential conductance measured at low
temperature will become broader. 

In Fig.3, differential conductance is calculated numerically by using
Eq.(\ref{dI/dV semi-classical}). By making comparisons between the
calculated results, the inclusion of the Coulomb interaction further
splits the conductance peaks. This splitting is expectable, because
the DOS of the QD is changed due to the electron-electron repulsion
in case of double occupation. However, the calculation also revivals
that all peaks become lower and slightly broaden when the PID effect
was included.

In this section, one of the potential observable phenomena due to
the PID effects is explored. We investigate the transport current
through a multi-terminals QPC device. The PID effect enters into the
problem since one of those terminals is assumed to satisfy the SGR
condition, consequently its role is included effectively according
to the adiabatic elimination approach presented in Sec.II. The steady
state current is calculated by combining the non-equilibrium Green's
function method with the semi-classical approach outlined in Sec.III.
The results show us that the height of differential conductance peaks
decreases as a result of the PID effect. This phenomena is controlled
by $\langle\langle\mathcal{R}_{\sigma}^{2}\rangle\rangle_{\mathrm{cl}}$,
which are in turn related to parameters of the proximity coupling,
e.g., $t_{0}$ and $\Delta$.

\section{Conclusion}

To conclude, we present an adiabatic elimination approach to incorporate
PE in SC-QD hybrid system. This method rely on the SGR condition,
where all QD levels are located inside the SC gap. Apart from those
already known PE induced pairing terms in the effective Hamiltonian
of the QD, we find new terms which are due to inter-electron interactions
in the QD. Those new terms represents the higher order couplings between
QD and the SC and can lead to dephasing of the QD in single electron
subspace. Using a semi-classical treatment, corresponding dephasing
time is studied and shown to be proportional to bulk SC gap in the
SGR in a special case. Physical implication of the PID effect is also
investigated based on a multi-terminals QPC model, results indicate
that differential conductance peaks would become lower and boarder
due to the presence of the PID effect at the zero temperature.
\begin{acknowledgments}
This work was supported by the National Natural Science Foundation
of China (Grant No.11121403) and the National 973 program (Grant No.2012CB922104
and No.2014CB921402).
\end{acknowledgments}
\appendix

\section{Full expression of $H_{sd}^{\mathrm{eff}}$}

Expanding Eq.(\ref{trans H}) up to the second order of $S$ as well
as $H_{sd}$
\begin{equation}
e^{-S}He^{S}\approx H_{s}+H_{d}+\frac{1}{2}[H_{sd},S],
\end{equation}
Notice that by chosen $S$ as shown in Eq.(\ref{S}), one has $[S,H_{s}+H_{d}]=H_{sd}$.
Therefore, terms that are linear in $H_{sd}$ does not appear in the
above expansion.

The interaction term $H_{sd}^{\mathrm{eff}}$ is contained in $[H_{sd},S]$.
By a direct calculation, it is written as
\begin{eqnarray}
H_{sd}^{\mathrm{eff}} & = & \frac{1}{2}\sum_{\mathbf{k},\mathbf{k}'}\{-\eta_{\mathbf{k}}a_{\uparrow}^{\dagger}a_{\downarrow}[\tilde{\alpha}_{\mathbf{k}'}\eta_{\mathbf{k'}}^{*}(\gamma_{k}\gamma_{-k'}^{\dagger}+\gamma_{-k}^{\dagger}\gamma_{k'})\nonumber \\
 &  & +\tilde{\beta}_{\mathbf{k}'}\lambda_{\mathbf{k'}}^{*}(\gamma_{k'}\gamma_{k}+\gamma_{-k}^{\dagger}\gamma_{-k'}^{\dagger})]\nonumber \\
 &  & +\lambda_{\mathbf{k}}a_{\uparrow}^{\dagger}a_{\downarrow}[\tilde{\alpha}_{\mathbf{k}'}\eta_{\mathbf{k'}}^{*}(\gamma_{-k'}^{\dagger}\gamma_{-k}^{\dagger}+\gamma_{k}\gamma_{k'})\nonumber \\
 &  & +\tilde{\beta}_{\mathbf{k}'}\lambda_{\mathbf{k'}}^{*}(\gamma_{k'}\gamma_{-k}^{\dagger}-\gamma_{k}\gamma_{-k'}^{\dagger})]\nonumber \\
 &  & -\eta_{\mathbf{k}}^{*}[a_{\uparrow}a_{\downarrow}(\tilde{\alpha}_{\mathbf{k}'}\eta_{\mathbf{k'}}\gamma_{k}^{\dagger}\gamma_{-k'}^{\dagger}-\tilde{\beta}_{\mathbf{k}'}\lambda_{\mathbf{k'}}\gamma_{k}^{\dagger}\gamma_{k'})\nonumber \\
 &  & +a_{\uparrow}^{\dagger}a_{\downarrow}^{\dagger}(\tilde{\alpha}_{\mathbf{k}'}\eta_{\mathbf{k'}}\gamma_{-k}\gamma_{k'}+\tilde{\beta}_{\mathbf{k}'}\lambda_{\mathbf{k'}}\gamma_{-k}\gamma_{-k'}^{\dagger})\nonumber \\
 &  & -n_{\uparrow}(\tilde{\alpha}_{\mathbf{k}'}\eta_{\mathbf{k'}}\gamma_{-k'}^{\dagger}\gamma_{-k}+\tilde{\beta}_{\mathbf{k}'}\lambda_{\mathbf{k'}}\gamma_{k'}\gamma_{-k})\\
 &  & +n_{\downarrow}(\tilde{\beta}_{\mathbf{k}'}\lambda_{\mathbf{k'}}\gamma_{-k'}^{\dagger}\gamma_{k}^{\dagger}-\tilde{\alpha}_{\mathbf{k}'}\eta_{\mathbf{k'}}\gamma_{k}^{\dagger}\gamma_{k'})]\\
 &  & +\lambda_{\mathbf{k}}^{*}[a_{\uparrow}a_{\downarrow}(\tilde{\alpha}_{\mathbf{k}'}\eta_{\mathbf{k'}}\gamma_{-k'}^{\dagger}\gamma_{-k}-\tilde{\beta}_{\mathbf{k}'}\lambda_{\mathbf{k'}}\gamma_{-k}\gamma_{k'})\nonumber \\
 &  & +a_{\uparrow}^{\dagger}a_{\downarrow}^{\dagger}(\tilde{\beta}_{\mathbf{k}'}\lambda_{\mathbf{k'}}\gamma_{k}^{\dagger}\gamma_{-k'}^{\dagger}-\tilde{\alpha}_{\mathbf{k}'}\eta_{\mathbf{k'}}\gamma_{k}^{\dagger}\gamma_{k'})\nonumber \\
 &  & +n_{\uparrow}(\tilde{\beta}_{\mathbf{k}'}\lambda_{\mathbf{k'}}\gamma_{k'}\gamma_{k}^{\dagger}-\tilde{\alpha}_{\mathbf{k}'}\eta_{\mathbf{k'}}\gamma_{k}^{\dagger}\gamma_{-k'}^{\dagger})\\
 &  & -n_{\downarrow}(\tilde{\alpha}_{\mathbf{k}'}\eta_{\mathbf{k'}}\gamma_{-k}\gamma_{k'}+\tilde{\beta}_{\mathbf{k}'}\lambda_{\mathbf{k'}}\gamma_{-k}\gamma_{-k'}^{\dagger})]\}\\
 &  & +h.c.\nonumber 
\end{eqnarray}
The PID interaction terms are given by Eqs.(A2-A5) as well as their
Hermitian conjugations.

\section{Dephasing time in the semi-classical treatment}

We first proof a few relations among the mean value as well as the
covariance matrix elements of the random number $\mathcal{R}_{\sigma}$
. Then those relations are used to estimate dephasing time in a special
case.

It follows from Eqs.(\ref{R_uparrow},\ref{semi-classic app}) that
\begin{eqnarray}
\langle\mathcal{R}_{\uparrow}\rangle_{\mathrm{cl}} & = & \frac{1}{2}\sum_{\mathbf{k},\mathbf{k}'}\langle\tilde{\alpha}_{\mathbf{k}'}(\eta_{\mathbf{k}}^{*}\eta_{\mathbf{k}'}\gamma_{-k'}^{\dagger}\gamma_{-k}-\lambda_{\mathbf{k}}^{*}\eta_{\mathbf{k}'}\gamma_{k}^{\dagger}\gamma_{-k'}^{\dagger})\nonumber \\
 &  & +\tilde{\beta}_{\mathbf{k}'}(\lambda_{\mathbf{k}}^{*}\lambda_{\mathbf{k}'}\gamma_{k'}\gamma_{k}^{\dagger}+\eta_{\mathbf{k}}^{*}\lambda_{\mathbf{k}'}\gamma_{k'}\gamma_{-k})\rangle_{\mathrm{sc}}+c.c.\nonumber \\
 & = & \frac{1}{2}\sum_{\mathbf{k}}\tilde{\beta}_{\mathbf{k}}|\lambda_{\mathbf{k}}|^{2}+c.c.\nonumber \\
 & = & \sum_{\mathbf{k}}\tilde{\beta}_{\mathbf{k}}|\lambda_{\mathbf{k}}|^{2}.
\end{eqnarray}
On the other hand, Eq.(\ref{R_downarrow}) implies that
\begin{eqnarray}
\langle\mathcal{R}_{\downarrow}\rangle_{\mathrm{cl}} & = & \frac{1}{2}\sum_{\mathbf{k},\mathbf{k}'}\langle-\tilde{\beta}_{\mathbf{k}'}(\lambda_{\mathbf{k}}^{*}\lambda_{\mathbf{k}'}\gamma_{-k}\gamma_{-k'}^{\dagger}+\eta_{\mathbf{k}}^{*}\lambda_{\mathbf{k}'}\gamma_{-k'}^{\dagger}\gamma_{k}^{\dagger})\nonumber \\
 &  & +\tilde{\alpha}_{\mathbf{k}'}(\eta_{\mathbf{k}}^{*}\eta_{\mathbf{k}'}\gamma_{k}^{\dagger}\gamma_{k'}-\lambda_{\mathbf{k}}^{*}\eta_{\mathbf{k}'}\gamma_{-k}\gamma_{k'})\rangle_{\mathrm{sc}}+c.c.\nonumber \\
 & = & -\sum_{\mathbf{k}}\tilde{\beta}_{\mathbf{k}}|\lambda_{\mathbf{k}}|^{2}.
\end{eqnarray}

Compare Eq.(B2) with Eq.(B1), then the following relation can be written
down, i.e., 
\begin{equation}
\langle\mathcal{R}_{\sigma}+\mathcal{R}_{\bar{\sigma}}\rangle_{\mathrm{cl}}=0.
\end{equation}

If we consider a special case such that $\varepsilon=-U/2$, then
some relations for covariance matrix elements can be further derived.
First, one can calculate that
\begin{eqnarray}
\langle\mathcal{R}_{\uparrow}^{2}\rangle_{\mathrm{cl}} & = & (\sum_{\mathbf{k}}\tilde{\beta}_{\mathbf{k}}|\lambda_{\mathbf{k}}|^{2})^{2}+\frac{1}{2}\sum_{\mathbf{k}}\tilde{\alpha}_{\mathbf{k}}|\eta_{\mathbf{k}}|^{2}\sum_{\mathbf{k}'}\tilde{\beta}_{\mathbf{k}}|\lambda_{\mathbf{k}}|^{2}\nonumber \\
 &  & +\frac{1}{4}\sum_{\mathbf{k},\mathbf{k}'}|\eta_{\mathbf{k}}|^{2}|\lambda_{\mathbf{k}'}|^{2}(\tilde{\alpha}_{\mathbf{k}}+\tilde{\beta}_{\mathbf{k}'}),
\end{eqnarray}
\begin{eqnarray}
\langle\mathcal{R}_{\downarrow}^{2}\rangle_{\mathrm{cl}} & = & (\sum_{\mathbf{k}}\tilde{\beta}_{\mathbf{k}}|\lambda_{\mathbf{k}}|^{2})^{2}-\frac{1}{2}\sum_{\mathbf{k}}\tilde{\alpha}_{\mathbf{k}}|\eta_{\mathbf{k}}|^{2}\sum_{\mathbf{k}'}\tilde{\beta}_{\mathbf{k}}|\lambda_{\mathbf{k}}|^{2}\nonumber \\
 &  & +\frac{1}{4}\sum_{\mathbf{k},\mathbf{k}'}|\eta_{\mathbf{k}}|^{2}|\lambda_{\mathbf{k}'}|^{2}(\tilde{\alpha}_{\mathbf{k}}+\tilde{\beta}_{\mathbf{k}'}),
\end{eqnarray}
and
\begin{eqnarray}
\langle\mathcal{R}_{\sigma}\mathcal{R}_{\bar{\sigma}}\rangle_{\mathrm{cl}} & = & -(\sum_{\mathbf{k}}\tilde{\beta}_{\mathbf{k}}|\lambda_{\mathbf{k}}|^{2})^{2}+\frac{1}{4}[(\sum_{\mathbf{k}}|\eta_{\mathbf{k}}\lambda_{\mathbf{k}}|\tilde{\alpha}_{\mathbf{k}})^{2}\nonumber \\
 &  & -(\sum_{\mathbf{k}}|\eta_{\mathbf{k}}\lambda_{\mathbf{k}}|\tilde{\beta}_{\mathbf{k}})^{2}].
\end{eqnarray}

Furthermore, when $\varepsilon=-U/2$
\begin{equation}
\sum_{\mathbf{k}}\tilde{\beta}_{\mathbf{k}}|\eta_{\mathbf{k}}|^{2}=\sum_{\mathbf{k}}\tilde{\alpha}_{\mathbf{k}}|\lambda_{\mathbf{k}}|^{2}.
\end{equation}

This is shown by first noticing that $\tilde{\alpha}_{\mathbf{k}}=\tilde{\beta}_{\mathbf{k}}$
from Eq.(\ref{para R_sig}). Then by converting the summation over
$\mathbf{k}$ to the integration over the kinetic energy of free electrons
$\xi$, one has 
\begin{eqnarray}
\sum_{\mathbf{k}}\tilde{\beta}_{\mathbf{k}}|\eta_{\mathbf{k}}|^{2} & = & t_{0}^{2}\sum_{\mathbf{k}}\frac{2\varepsilon\cos^{2}\theta_{k}}{E_{\mathbf{k}}^{2}-\varepsilon^{2}}\nonumber \\
 & = & \varepsilon t_{0}^{2}\int_{-\infty}^{\infty}d\xi\frac{N_{0}}{\xi^{2}+\Delta^{2}-\varepsilon^{2}}(1+\frac{\xi}{\sqrt{\xi^{2}+\Delta^{2}}})\nonumber \\
 & = & 2\varepsilon t_{0}^{2}\int_{0}^{\infty}d\xi\frac{N_{0}}{\xi^{2}+\Delta^{2}-\varepsilon^{2}},
\end{eqnarray}
and
\begin{eqnarray}
\sum_{\mathbf{k}}\tilde{\alpha}_{\mathbf{k}}|\eta_{\mathbf{k}}|^{2} & = & t_{0}^{2}\sum_{\mathbf{k}}\frac{2\varepsilon\sin^{2}\theta_{k}}{E_{\mathbf{k}}^{2}-\varepsilon^{2}}\nonumber \\
 & = & \varepsilon t_{0}^{2}\int_{-\infty}^{\infty}d\xi\frac{N_{0}}{\xi^{2}+\Delta^{2}-\varepsilon^{2}}(1-\frac{\xi}{\sqrt{\xi^{2}+\Delta^{2}}})\nonumber \\
 & = & 2\varepsilon t_{0}^{2}\int_{0}^{\infty}d\xi\frac{N_{0}}{\xi^{2}+\Delta^{2}-\varepsilon^{2}}.
\end{eqnarray}
Notice that the lower integration bound is pushed to $-\infty$, because
the contribution from the integrand approaches to zero for $\xi$
that are deeply inside the Fermi-surface. Compare Eq.(B8) and Eq.(B9),
thus Eq.(B7) follows directly. With Eq.(B7), Eqs.(B4-B6) implies that
\begin{equation}
\langle\langle\mathcal{R}_{\downarrow}^{2}\rangle\rangle_{\mathrm{cl}}=\langle\langle\mathcal{R}_{\uparrow}^{2}\rangle\rangle_{\mathrm{cl}}-\langle\mathcal{R}_{\uparrow}\rangle_{\mathrm{cl}}^{2},\mbox{ }\langle\langle\mathcal{R}_{\sigma}\mathcal{R}_{\bar{\sigma}}\rangle\rangle_{\mathrm{cl}}=0.
\end{equation}

Now consider the estimation of the dephasing time. With the help of
Eqs.(B3,B10), $\Gamma$ in Eq.(\ref{decoherence factor sc app para})
is written as follows
\begin{equation}
\Gamma=\sqrt{\langle\langle\mathcal{R}_{\uparrow}^{2}\rangle\rangle_{\mathrm{cl}}-\frac{1}{8}\Omega^{2}}.
\end{equation}

Using the integration substitution procedure given in Eqs.(B8-B9),
it is shown that
\begin{equation}
\langle\mathcal{R}_{\uparrow}\rangle_{\mathrm{cl}}=\frac{2\pi\varepsilon N_{0}t_{0}^{2}}{\sqrt{\Delta^{2}-\varepsilon^{2}}},
\end{equation}
and
\begin{equation}
\langle\langle\mathcal{R}_{\uparrow}^{2}\rangle\rangle_{\mathrm{cl}}=\frac{1}{2}\langle\mathcal{R}_{\uparrow}\rangle_{\mathrm{cl}}^{2}+4\pi t_{0}^{4}\varepsilon^{2}N_{0}^{2}\frac{\sqrt{\Lambda_{D}^{2}-\Delta^{2}}}{(\Delta^{2}-\varepsilon^{2})^{\frac{3}{2}}}.
\end{equation}
Here, the energy cut-off $\Lambda_{D}$ has been introduced in Sec.
III. Then dephasing time shown in Eq.(\ref{dephasing time}) is obtained
by substituting Eqs.(B12,B13) into Eq.(B11) and noticing $\Delta_{d}$
introduced in Eq.(\ref{Dd}).

\section{Derivation of the QD Green's function}

Consider the non-equilibrium counterpart of the QD Green's function
\begin{equation}
g_{\sigma}(t,t')=-i\langle Td_{\sigma}(t)d_{\sigma}^{\dagger}(t')\rangle
\end{equation}
where $T$ is time ordering operator on the closed time-path contour.

Take time derivative with respect to $t$, 
\begin{eqnarray}
i\partial_{t}g_{\sigma}(t,t') & = & \delta(t-t')+(\varepsilon+\mathcal{R}_{\sigma})g_{\sigma}(t,t')+Ug_{\sigma}^{(2)}(t,t')\nonumber \\
 &  & \pm\Delta_{d}F_{\bar{\sigma}\sigma}(t,t')+\sum_{\nu,\mathbf{k}}t_{\mathbf{k}}^{(\nu)}G_{\mathbf{k}\sigma}^{(\nu)}(t,t').\label{RP_92_122}
\end{eqnarray}
Here, $\Delta_{d}$ and $-\Delta_{d}$ in front of $F_{\bar{\sigma}\sigma}(t,t')$
correspond to taking $\sigma=\uparrow$ and $\downarrow$, respectively.
New Green's functions have been introduced in above equation. Their
definitions are listed in Tab. I.

\begin{table}
\caption{Definitions for various Green's functions using in Appendix C. The
sub-index means atomic limit (AL) \cite{STM prob local density state},
where $t_{0}$ and $t_{1}$ are set to zero. Here, $\nu\in\{L,R\}$.}

\centering{}%
\begin{tabular}{c}
\hline 
$F_{\bar{\sigma}\sigma}(t,t')=-i\left\langle Td_{\bar{\sigma}}^{\dagger}(t)d_{\sigma}^{\dagger}(t')\right\rangle $\tabularnewline
$F_{\bar{\sigma}\sigma}^{(2)}(t,t')=-i\left\langle T[n_{\sigma}d_{\bar{\sigma}}^{\dagger}](t)d_{\sigma}^{\dagger}(t')\right\rangle $\tabularnewline
$F_{\mathbf{k}\sigma}^{(\nu)}(t,t')=-i\left\langle Tc_{\nu\mathbf{k}\bar{\sigma}}^{\dagger}(t)d_{\sigma}^{\dagger}(t')\right\rangle $\tabularnewline
$g_{\sigma}(t,t')=-i\langle Td_{\sigma}(t)d_{\sigma}^{\dagger}(t')\rangle$\tabularnewline
$g_{\sigma}^{(2)}(t,t')=-i\left\langle T[n_{\bar{\sigma}}d_{\sigma}](t)d_{\sigma}^{\dagger}(t')\right\rangle $\tabularnewline
$G_{\mathbf{k}\sigma}^{(\nu)}(t,t')=-i\left\langle Tc_{\nu\mathbf{k}\sigma}(t)d_{\sigma}^{\dagger}(t')\right\rangle $\tabularnewline
\hline 
$g_{\sigma}^{(e)}(t,t')=-i\left\langle Td_{\sigma}(t)d_{\sigma}^{\dagger}(t')\right\rangle _{0}$\tabularnewline
$\mbox{ }g_{\sigma}^{(h)}(t,t')=-i\left\langle Td_{\sigma}^{\dagger}(t)d_{\sigma}(t')\right\rangle _{0}$\tabularnewline
$G_{\mathbf{k}\sigma}^{(\nu,e)}(t,t')=-i\left\langle Tc_{\nu\mathbf{k}\sigma}(t)c_{\nu\mathbf{k}\sigma}^{\dagger}(t')\right\rangle _{0}$\tabularnewline
$\mbox{ }G_{\mathbf{k}\sigma}^{(\nu,h)}(t,t')=-i\left\langle Tc_{\nu\mathbf{k}\sigma}^{\dagger}(t)c_{\nu\mathbf{k}\sigma}(t')\right\rangle _{0}$\tabularnewline
$g_{\sigma}^{(2,e)}(t,t')=-i\left\langle T[n_{\bar{\sigma}}d_{\sigma}](t)d_{\sigma}^{\dagger}(t')\right\rangle _{0}$\tabularnewline
$g_{\sigma}^{(2,h)}(t,t')=-i\left\langle T[n_{\bar{\sigma}}d_{\sigma}^{\dagger}](t)d_{\sigma}(t')\right\rangle _{0}$\tabularnewline
\hline 
$\Sigma_{\sigma}^{(e/h)}(t,t')=\sum_{\nu,\mathbf{k}}|t_{\mathbf{k}}^{(\nu)}|^{2}G_{\mathbf{k}\sigma}^{(\nu,e/h)}(t,t')$\tabularnewline
\hline 
\end{tabular}
\end{table}
Motion equation of $F_{\bar{\sigma}\sigma}(t,t')$ is given by
\begin{eqnarray}
i\partial_{t}F_{\bar{\sigma}\sigma}(t,t') & = & -(\varepsilon+\mathcal{R}_{\bar{\sigma}})F_{\bar{\sigma}\sigma}(t,t')-UF_{\bar{\sigma}\sigma}^{(2)}(t,t')\nonumber \\
 &  & \pm\Delta_{d}^{*}g_{\sigma}(t,t')-\sum_{\nu,\mathbf{k}}t_{\mathbf{k}}^{(\nu)*}F_{\mathbf{k}\sigma}^{(\nu)}(t,t').\label{RP_92_123}
\end{eqnarray}

Due to the Coulomb interaction, two-particle Green's functions $g_{\sigma}^{(2)}(t,t')$
and $F_{\bar{\sigma}\sigma}^{(2)}(t,t')$ appear in above equations.
Motion equations of those two Green's functions are calculated as
\begin{eqnarray}
i\partial_{t}g_{\sigma}^{(2)}(t,t') & = & \delta(t-t')\left\langle n_{\bar{\sigma}}(t)\right\rangle +(\varepsilon+U+\mathcal{R}_{\sigma})g_{\sigma}^{(2)}(t,t')\nonumber \\
 &  & \mp i\Delta_{d}\left\langle T[d_{\sigma}d_{\bar{\sigma}}^{\dagger}d_{\sigma}^{\dagger}](t)d_{\sigma}^{\dagger}(t')\right\rangle \nonumber \\
 &  & +\sum_{\nu,\mathbf{k}}\{t_{\mathbf{k}}^{(\nu)*}i\left\langle T(d_{\bar{\sigma}}d_{\sigma}c_{\nu\mathbf{k}\bar{\sigma}}^{\dagger})(t)d_{\sigma}^{\dagger}(t')\right\rangle \nonumber \\
 &  & +t_{\mathbf{k}}^{(\nu)}[-i\left\langle T(n_{\bar{\sigma}}c_{\nu\mathbf{k}\sigma})(t)d_{\sigma}^{\dagger}(t')\right\rangle \nonumber \\
 &  & +i\left\langle T(d_{\bar{\sigma}}^{\dagger}d_{\sigma}c_{\nu\mathbf{k}\bar{\sigma}})(t)d_{\sigma}^{\dagger}(t')\right\rangle ]\},
\end{eqnarray}
as well as
\begin{eqnarray}
 &  & i\partial_{t}F_{\bar{\sigma}\sigma}^{(2)}(t,t')\nonumber \\
 & = & \delta(t-t')\left\langle [d_{\bar{\sigma}}^{\dagger}d_{\sigma}^{\dagger}](t)\right\rangle -(\varepsilon+U+\mathcal{R}_{\bar{\sigma}})F_{\bar{\sigma}\sigma}^{(2)}(t,t')\nonumber \\
 &  & \mp i\Delta_{d}^{*}\left\langle T[d_{\bar{\sigma}}d_{\sigma}d_{\bar{\sigma}}^{\dagger}](t)d_{\sigma}^{\dagger}(t')\right\rangle \nonumber \\
 &  & +\sum_{\nu,\mathbf{k}}\{it_{\mathbf{k}}^{(\nu)}\left\langle T[c_{\nu\mathbf{k}\sigma}d_{\bar{\sigma}}^{\dagger}d_{\sigma}^{\dagger}](t)d_{\sigma}^{\dagger}(t')\right\rangle \nonumber \\
 &  & +t_{\mathbf{k}}^{(\nu)*}[i\left\langle T(c_{\nu\mathbf{k}\bar{\sigma}}^{\dagger}n_{\sigma})(t)d_{\sigma}^{\dagger}(t')\right\rangle \nonumber \\
 &  & -i\left\langle T(c_{\nu\mathbf{k}\sigma}^{\dagger}d_{\bar{\sigma}}^{\dagger}d_{\sigma})(t)d_{\sigma}^{\dagger}(t')\right\rangle ]\}.
\end{eqnarray}

To close motion equations at the two-particle level, we employ the
mean-field truncation scheme \cite{MF app 1,MF app 2}. In this scheme,
two equal time operators in new Green's functions that appeared in
rhs of Eqs.(C4,C5) are paired up to form an averaging value. Then
those new Green's functions are rewritten as sums of all possible
pairings multiplied by one-particle Green's functions, which are already
introduced. Finally the averaging value of all equal time pairs are
replaced by averaging under the AL. For example,
\begin{eqnarray}
 &  & i\left\langle T(d_{\bar{\sigma}}d_{\sigma}c_{\nu\mathbf{k}\bar{\sigma}}^{\dagger})(t)d_{\sigma}^{\dagger}(t')\right\rangle \nonumber \\
 & \approx & -\left\langle d_{\bar{\sigma}}(t)d_{\sigma}(t)\right\rangle _{0}F_{\mathbf{k}\sigma}^{(\nu)}(t,t')-\left\langle c_{\nu\mathbf{k}\bar{\sigma}}^{\dagger}(t)d_{\bar{\sigma}}(t)\right\rangle _{0}g_{\sigma}(t,t')\nonumber \\
 &  & -i\left\langle c_{\nu\mathbf{k}\bar{\sigma}}^{\dagger}(t)d_{\sigma}(t)\right\rangle _{0}\left\langle Td_{\bar{\sigma}}(t)d_{\sigma}^{\dagger}(t')\right\rangle .
\end{eqnarray}

Since the averaging is taken under the AL, thus terms like $\langle d_{\bar{\sigma}}(t)d_{\sigma}(t)\rangle_{0}$
as well as the QD-lead cross terms like $\langle c_{\nu\mathbf{k}\sigma}(t)d_{\sigma'}^{\dagger}(t')\rangle_{0}$
or $\langle c_{\nu\mathbf{k}\sigma}(t)d_{\sigma'}(t')\rangle_{0}$
will all vanish. Therefore, Eqs.(C4,C5) are rewritten as
\begin{eqnarray}
 &  & (i\partial_{t}-\varepsilon-U)g_{\sigma}^{(2)}(t,t')\nonumber \\
 & = & \left\langle n_{\bar{\sigma}}(t)\right\rangle _{0}[\delta(t-t')+\mathcal{R}_{\sigma}g_{\sigma}(t,t')]\pm\Delta_{d}\left\langle n_{\sigma}(t)\right\rangle _{0}F_{\bar{\sigma}\sigma}(t,t')\nonumber \\
 &  & +\sum_{\nu,\mathbf{k}}t_{\mathbf{k}}^{(\nu)}\left\langle n_{\bar{\sigma}}(t)\right\rangle _{0}G_{\mathbf{k}\sigma}^{(\nu)}(t,t'),\label{RP_92_149}
\end{eqnarray}
and
\begin{eqnarray}
 &  & (i\partial_{t}+\varepsilon+U)F_{\bar{\sigma}\sigma}^{(2)}(t,t')\nonumber \\
 & = & -\left\langle n_{\sigma}(t)\right\rangle _{0}[\mathcal{R}_{\bar{\sigma}}F_{\bar{\sigma}\sigma}(t,t')+\sum_{\nu,\mathbf{k}}t_{\mathbf{k}}^{(\nu)*}F_{\mathbf{k}\sigma}^{(\nu)}(t,t')]\nonumber \\
 &  & \pm\Delta_{d}^{*}\left\langle n_{\bar{\sigma}}(t)\right\rangle _{0}g_{\sigma}(t,t').
\end{eqnarray}

Eqs.(C2,C3,C7,C8) together with the motion equations of $G_{\mathbf{k}\sigma}^{(\nu)}(t,t')$
and $F_{\mathbf{k}\sigma}^{(\nu)}(t,t')$ now constitute a closed
set of equations. By inverting differential operators, this set of
equations are rewritten in the following integration form
\begin{eqnarray}
g_{\sigma} & = & g_{\sigma}^{(e)}+\mathcal{R}_{\sigma}[g_{\sigma}^{(e)}*g_{\sigma}]\pm\Delta_{d}[g_{\sigma}^{(e)}*F_{\bar{\sigma}\sigma}]+U[g_{\sigma}^{(e)}*g_{\sigma}^{(2)}]\nonumber \\
 &  & +[g_{\sigma}^{(e)}*\Sigma_{\sigma}^{(e)}*g_{\sigma}],\label{RP_92_151}
\end{eqnarray}
\begin{eqnarray}
g_{\sigma}^{(2)} & = & g_{\sigma}^{(2,e)}+\mathcal{R}_{\sigma}[g_{\sigma}^{(2,e)}*g_{\sigma}]\pm\Delta_{d}[g_{\bar{\sigma}}^{(2,e)}*F_{\bar{\sigma}\sigma}]\nonumber \\
 &  & +[g_{\sigma}^{(2,e)}*\Sigma_{\sigma}^{(e)}*g_{\sigma}],\label{RP_92_153}
\end{eqnarray}
\begin{eqnarray}
F_{\bar{\sigma}\sigma} & = & \pm\Delta_{d}^{*}[g_{\sigma}^{(h)}*g_{\sigma}]-U[g_{\sigma}^{(h)}*F_{\bar{\sigma}\sigma}^{(2)}]-\mathcal{R}_{\bar{\sigma}}[g_{\sigma}^{(h)}*F_{\bar{\sigma}\sigma}]\nonumber \\
 &  & +[g_{\sigma}^{(h)}*\Sigma_{\sigma}^{(h)}*F_{\bar{\sigma}\sigma}],\label{RP_92_152}
\end{eqnarray}
and
\begin{eqnarray}
F_{\bar{\sigma}\sigma}^{(2)} & = & \pm\Delta_{d}^{*}[g_{\sigma}^{(2,h)}*g_{\sigma}]-\mathcal{R}_{\bar{\sigma}}[g_{\bar{\sigma}}^{(2,h)}*F_{\bar{\sigma}\sigma}]\nonumber \\
 &  & +[g_{\bar{\sigma}}^{(2,h)}*\Sigma_{\sigma}^{(h)}*F_{\bar{\sigma}\sigma}],
\end{eqnarray}
where $f*g=\int_{c}dt''f(t,t'')g(t'',t')$ means a time convolution
on the time contour $c$. New Green's functions as well as self energies
(see Tab. I) are defined under the AL, thus can be evaluated exactly
without introducing further approximations.

Applying Langreth identity \cite{Langreth}, Eqs.(C9-C12) are analytically
continued to give equations for retarded Green's function of the QD.
Then $g_{\sigma}^{r}(\omega)$ is solved by performing Fourier transform
on the corresponding set of equations.

\section{Cumulant expansion of $\langle g_{\sigma}^{r}(\omega)\rangle_{\mathrm{cl}}$}

When $\varepsilon=-U/2$ and $\langle n_{\sigma}\rangle_{0}=1/2$,
$g_{\sigma}^{r}(\omega)$ is written as follows by using Eq.(\ref{g_retard mean-field}),
i.e.,
\begin{equation}
g_{\sigma}^{r}(\omega)=\frac{\omega}{\omega^{2}-\varepsilon^{2}+(i\Gamma_{0}-\mathcal{R}_{\sigma})\omega-\frac{\Delta_{d}^{2}\omega^{2}}{\omega^{2}-\varepsilon^{2}+(i\Gamma_{0}+\mathcal{R}_{\bar{\sigma}})\omega}}.
\end{equation}

To evaluate $\langle g_{\sigma}^{r}(\omega)\rangle_{\mathrm{cl}}$,
we propose following cumulant expansion ansatz
\begin{equation}
\langle g_{\sigma}^{r}(\omega)\rangle_{\mathrm{cl}}=\frac{\omega}{\mathcal{A}(\omega)+f_{\sigma}^{(1)}+f_{\sigma}^{(2)}+...-\frac{\Delta_{d}^{2}\omega^{2}}{\mathcal{A}(\omega)+h_{\sigma}^{(1)}+h_{\sigma}^{(2)}+...}},
\end{equation}
where $\mathcal{A}(\omega)=\omega^{2}-\varepsilon^{2}+i\Gamma_{0}\omega$.
$f_{\sigma}^{(j)}$ and $h_{\sigma}^{(j)}$ are cumulants that are
j-th order in $\mathcal{R}_{\uparrow}$ or $\mathcal{R}_{\downarrow}$.

Assuming that both $\mathcal{R}_{\sigma}$ as well as the cumulants
are small quantities, then by comparing the expansions of Eq.(D2)
as well as the ansatz up to the second order, following equations
are found
\begin{equation}
\chi h_{\sigma}^{(1)}+f_{\sigma}^{(1)}=\chi\omega\langle\mathcal{R}_{\bar{\sigma}}\rangle_{\mathrm{cl}}-\omega\langle\mathcal{R}_{\sigma}\rangle_{\mathrm{cl}},
\end{equation}
and
\begin{eqnarray}
 &  & \chi h_{\sigma}^{(2)}-\chi\mathcal{A}^{-1}(h_{\sigma}^{(1)})^{2}+f_{\sigma}^{(2)}-\mathcal{B}^{-1}(\chi h_{\sigma}^{(1)}+f_{\sigma}^{(1)})^{2}\nonumber \\
 & = & -\chi\mathcal{A}^{-1}\omega^{2}\langle\mathcal{R}_{\bar{\sigma}}^{2}\rangle_{\mathrm{cl}}-\mathcal{B}^{-1}\langle(\chi\omega\mathcal{R}_{\bar{\sigma}}-\omega\mathcal{R}_{\sigma})^{2}\rangle_{\mathrm{cl}},
\end{eqnarray}
where $\chi=\omega^{2}\Delta_{d}^{2}\mathcal{A}^{-2}$ and $\mathcal{B}=\mathcal{A}(1-\chi)$. 

From Eqs.(D3,D4), the cumulants are solved as
\begin{equation}
\mbox{ }f_{\sigma}^{(1)}=h_{\sigma}^{(1)}=-\omega\langle\mathcal{R}_{\sigma}\rangle_{\mathrm{cl}},
\end{equation}
as well as
\begin{eqnarray}
\mbox{ }f_{\sigma}^{(2)}=h_{\bar{\sigma}}^{(2)} & = & -\frac{\omega}{\mathcal{A}-\Delta_{d}^{2}\omega^{2}\mathcal{A}^{-1}}\omega\langle\langle\mathcal{R}_{\sigma}^{2}\rangle\rangle_{\mathrm{cl}}\nonumber \\
 & = & -g_{\sigma}^{r}(\omega)|_{\mathcal{R}_{\sigma}\equiv0}\omega\langle\langle\mathcal{R}_{\sigma}^{2}\rangle\rangle_{\mathrm{cl}}.
\end{eqnarray}
Here, Eq.(B3) has been used in derivation. 

Therefore,
\begin{eqnarray}
 &  & f_{\sigma}^{(1)}+f_{\sigma}^{(2)}+i\Gamma_{0}\omega\nonumber \\
 & = & \{-\omega\langle\mathcal{R}_{\sigma}\rangle_{\mathrm{cl}}-\omega\langle\langle\mathcal{R}_{\sigma}^{2}\rangle\rangle_{\mathrm{cl}}\mathrm{Re}[g_{\sigma}^{r}(\omega)|_{\mathcal{R}_{\sigma}\equiv0}]\}\nonumber \\
 &  & +i\{\Gamma_{0}-\langle\langle\mathcal{R}_{\sigma}^{2}\rangle\rangle_{\mathrm{cl}}\mathrm{Im}[g_{\sigma}^{r}(\omega)|_{\mathcal{R}_{\sigma}\equiv0}]\omega\}\nonumber \\
 & = & q_{\sigma}^{(e)}(\omega)+i\Gamma_{\sigma}\omega,
\end{eqnarray}
and
\begin{eqnarray}
 &  & h_{\sigma}^{(1)}+h_{\sigma}^{(2)}+i\Gamma_{0}\omega\nonumber \\
 & = & \{\omega\langle\mathcal{R}_{\bar{\sigma}}\rangle_{\mathrm{cl}}-\omega\langle\langle\mathcal{R}_{\bar{\sigma}}^{2}\rangle\rangle_{\mathrm{cl}}\mathrm{Re}[g_{\bar{\sigma}}^{r}(\omega)|_{\mathcal{R}_{\bar{\sigma}}\equiv0}]\}\nonumber \\
 &  & +i\{\Gamma_{0}-\langle\langle\mathcal{R}_{\bar{\sigma}}^{2}\rangle\rangle_{\mathrm{cl}}\mathrm{Im}[g_{\bar{\sigma}}^{r}(\omega)|_{\mathcal{R}_{\bar{\sigma}}\equiv0}]\omega\}\nonumber \\
 & = & q_{\sigma}^{(h)}(\omega)+i\Gamma_{\bar{\sigma}}\omega.
\end{eqnarray}

$\Gamma_{\sigma}$ as shown in Eq.(\ref{Gamma_sig corrected}) can
then be derived using Eq.(\ref{dI/dV}) to write 
\begin{equation}
\mathrm{Im}[g_{\sigma}^{r}(eV)|_{\mathcal{R}_{\sigma}\equiv0}]=-\frac{h}{2\Gamma_{0}e^{2}}\frac{dI_{\mathrm{tot}}}{dV}|_{\mathcal{R}_{\sigma}\equiv0}.
\end{equation}

\end{document}